\documentclass[11pt]{article}
\pdfoutput=1
\usepackage[utf8]{inputenc}
\usepackage{multirow}
\usepackage{amsmath, amsfonts, amssymb}
\usepackage{comment}
\usepackage{graphicx}
\usepackage{psfrag}
\usepackage{amsthm}
\usepackage[usenames,dvipsnames,svgnames,table]{xcolor}
\usepackage{enumerate}
\usepackage{arydshln}
\usepackage{soul}
 \usepackage{slashed}
\usepackage{subcaption}
\usepackage{mathrsfs}
\usepackage{ytableau}
 \usepackage{a4wide}
  \usepackage{tikz}
  \usepackage{tikz-cd}
  \usetikzlibrary{shapes.geometric}
  \usepackage{tcolorbox}
  \usepackage[usenames, dvipsnames]{xcolor}
  \definecolor{dark-gray}{gray}{0.20}
  \definecolor{gray}{gray}{0.30}
  \definecolor{light-gray}{gray}{0.80}
  \definecolor{dark-red}{rgb}{0.7,0,0}
  \definecolor{dark-green}{rgb}{0.1,0.4,0}
  \definecolor{dark-blue}{rgb}{0.3,0.3,0.7}
  \definecolor{light-blue}{rgb}{0.8,0.8,1}
      \definecolor{swamp}{RGB}{240, 199, 197}
      
  \usepackage{pifont}

\usepackage{newunicodechar} 
\usepackage{setspace}

\newcommand{\be}{\begin{equation}}
\newcommand{\ee}{\end{equation}}
\newcommand{\eq}[1]{(\ref{#1})}
\newcommand{\ket}[1]{\vert #1 \rangle}

\def\be{\begin{equation}}
\def\ee{\end{equation}}
\def\bea{\begin{eqnarray}}
\def\eea{\end{eqnarray}}

\newcommand{\al}{\alpha}

\newcommand{\mN}{\mathcal{N}}

\newcommand{\R}{\mathbb{R}}

\AtBeginDocument{

}

\newcommand{\Sp}{\mathit{Sp}}

\numberwithin{equation}{section}

\usepackage{jheppub}
\usepackage{hyperref}
\usepackage{cleveref}

\hypersetup{
	colorlinks=true,
	linkcolor=dark-blue,
	citecolor=dark-red,
	urlcolor=dark-green,
	linktoc=page
}

\theoremstyle{definition}

\theoremstyle{remark}

\crefname{appendix}{Appendix}{Appendices}

\title{\centering Exotic supergravities and the Swampland}

\author{Miguel Montero} \affiliation{$^3$ Instituto de F\'{i}sica Te\'{o}rica IFT-UAM/CSIC,
C/ Nicol\'{a}s Cabrera 13-15, Campus de Cantoblanco, 28049 Madrid, Spain}
\author{and Michelangelo Tartaglia}

\emailAdd{miguel.montero@csic.es}
\emailAdd{michelangelo.tartaglia@estudiante.uam.es}
\abstract{In six dimensions, there is an exotic $\mathcal{N}=(4,0)$ supermultiplet that contains only fields of spin $\leq 2$, but no graviton, and that on a circle reduces to 5D $\mathcal{N}=4$ supergravity. It has been proposed that, if suitable interactions exist, the $(4,0)$ theory might provide a consistent alternative UV completion for $\mathcal{N}=4$ 5D supergravity, realizing a supersymmetric version of asymptotic safety. In this note we argue that any Lorentz-invariant $(4,0)$ theory (interacting or not) carries an exact global symmetry when compactified on $S^1$, and is therefore incompatible with the Swampland no global symmetries conjecture. Another example of exotic supergravity, the 6D $(3,1)$ theory, does not have this problem. We study the general case and find that the only exotic spin-2 field that reduces to Einsteinian gravity and has no global symmetries when compactified on a high-dimensional torus is that of the $(3,1)$ theory. All other possibilities either yield several gravitons or have global symmetries.}

\begin{document}
\emergencystretch 3em
\hypersetup{pageanchor=false}
\makeatletter
\let\old@fpheader\@fpheader
\preprint{IFT-24-25}

\makeatother

\maketitle

\hypersetup{
    pdftitle={},
    pdfauthor={},
    pdfsubject={}
}

\newcommand{\remove}[1]{\textcolor{red}{\sout{#1}}}

\section{Introduction and summary}
Is String Theory, in some sense, the unique consistent quantum theory of gravity? This idea, commonly called ``string universality'' \cite{Adams:2010zy} or the ``String Lamppost Principle '' \cite{Montero:2020icj}, has received a lot of attention in recent years in the context of the Swampland Program \cite{Palti:2019pca,vanBeest:2021lhn}, which aims to identify those effective field theories that can never arise as the low-energy limit of consistent quantum gravity. In a sense, string universality is the ``endgame'' of the Swampland program -- universality is achieved when the list of EFT's allowed by Swampland principles exactly matches those which can be obtained from string compactifications.

Often, the general question of universality is restricted to a particular subset of EFT's -- such as, for instance, those in a given dimension and number of supercharges -- to make it manageable. In fact the question of string universality was first posed in \cite{Fiol:2008gn} in the context of 10d $\mathcal{N}=1$ supergravities, where the Green-Schwarz mechanism allows for just four possibilities for the gauge group of the theory (these being $E_8\times E_8$, $SO(32),$ $E_8\times U(1)^{248}$ and $U(1)^{496}$), of which only two ($E_8\times E_8$, $SO(32)$) are realized in String theory. The work \cite{Adams:2010zy} solved this question (which was addressed later from a Swampland viewpoint \cite{Kim:2019vuc}). Since then, string universality has been achieved in ungauged supergravity theories with sixteen supercharges both in nine and eight dimensions \cite{Montero:2020icj,Hamada:2021bbz,Bedroya:2021fbu}, is close to being attained in seven, and six dimensions seems within reach \cite{Baykara:2023plc}.

Given the considerable recent progress in theories with 16 supercharges, it may be surprising that universality with 32 supercharges has in fact \emph{not} been achieved! As is well-known, 32 is the largest amount of supercharges for which only fields of spin $\leq 2$ exist in a short multiplet. This amount of supersymmetry is minimal in 11d, where it corresponds to 11d supergravity, which is the low-energy limit of M-theory. In 10 dimensions, there are two such supergravities, IIA and IIB, realized as the low-energy limits of IIA and IIB strings respectively. Below dimension 10, there is a single Einsteinian (i.e. theory of a symmetric two-index tensor) supergravity with 32 supercharges, realized as the low energy limit of M-theory or type II strings on a torus $T^n$. However, these are not the only theories containing fields of spins $\leq 2$!

In six dimensions there is an exotic $(4,0)$ chiral multiplet \cite{Strathdee:1986jr,Hull:2000zn} which contains scalars and 2-forms, but no graviton. Instead, there is a fundamental four-index object $C_{MNPQ}$ with the symmetries of the Riemann tensor (so antisymmetric under $M\leftrightarrow N$ and $P\leftrightarrow Q$, but symmetric under $MN\leftrightarrow PQ$). Although it has many indices, this field only propagates degrees of freedom with spin less than two, in the physical sense that polarizations around any two-plane rotate with a phase $e^{2\pi i s \theta}$ with $s\leq 2$, so it does not count as ``higher spin''\footnote{More precisely, this field transforms in the $\mathbf{5}$ of the massless 6D little group, and this representation only has physical polarizations of spin $\leq 2$.}. More precisely, the various arguments that higher spin fields always come in infinite towers of increasing spin are really about having at least one physical polarization with spin $\geq 3 $, and do not obviously apply to fields like $C_{MNPQ}$ \cite{Alba:2015upa}. So the 6D $(4,0)$ multiplet can give, a priori, a reasonable theory with 32 supercharges and a finite number of fields. It is part of the string universality question with 32 supercharges. Furthermore, as shown in \cite{Hull:2000rr,Hull:2000zn}, the $(4,0)$ theory reduced on $S^1$ yields the maximal, Einsteinian $\mathcal{N}=4$ supergravity in five dimensions \footnote{Exotic representations of low spin of the Lorentz group exist only in $d\geq 6$ and they all reduce to Einsteinian theories when compactified to 5 or less dimensions, as we will explain in Section \ref{sec:2}.}. This maximal supergravity appears in the string landscape, as the low-energy limit of M-theory on $T^6$, which led \cite{Hull:2000rr,Hull:2000zn} to propose that the $(4,0)$ theory may actually be realized in the high-energy regime of a very stringy point in the moduli space of M-theory on $T^6$. We do not know yet if this is indeed the case; but if so, the $(4,0)$ theory would constitute a new phase of gravity and a counterexample to the Emergent String Conjecture of \cite{Lee:2019wij}, which posits that the only two phases of gravity with a semiclassical spacetime are perturbative strings or Einsteinian.

The $(4,0)$ theory does not appear in  the high-energy regime of any asymptotic corner of the moduli space of M-theory on $T^6$; these are are all known \cite{Obers:1998fb}. So if it is there, it can only appear in the deep UV of a so-far uncontrolled region of moduli space. In Einsteinian theories, the UV regime of a bulk moduli space point can be traded by an asymptotic limit, since the $S^1$ volume is a modulus and can be tuned to be very large. But in the dimensional reduction of the $(4,0)$ on $S^1$ there is no radion, so the fact that we have not seen this theory in asymptotic corners is completely consistent with the proposal. Related to this, the $(4,0)$ multiplet is actually conformal \cite{Hull:2002hmj,Hull:2022vlv}, which motivated \cite{Hull:2002hmj} to propose that there might be an interacting $(4,0)$ theory without any scale; the 5D Planck length would then be the compactification radius, which is the only scale in the theory\footnote{Agreement with 5D black hole spectrum then presumably requires some sort of nonlocal physics in 6D such as tensionless strings, etc.}. In this sense, the $(4,0)$ theory is a supersymmetric realization of the ``asymptotic safety'' scenario for quantum gravity \cite{Percacci:2007sz,Eichhorn:2018yfc}. The reader may have noticed the strong similarities between this story and the well established ``field theory asymptotic safety'' for maximal 5D Super Yang-Mills, which arises as the low-energy limit of the $(2,0)$ SCFT on a circle.

So, all in all, the $(4,0)$ proposal (as well as its $(3,1)$ cousin \cite{Hull:2002hmj}, which also has 32 supercharges) remains a hypothetical, yet very appealing proposal for an exotic phase of gravity. The goal of this note is to confront the $(4,0)$ proposal with the Swampland program. Although Swampland constraints may not apply directly to the $(4,0)$ theory, they do apply to its dimensional reduction on $S^1$, since it is an Einstein theory of gravity. We find that the $(4,0)$ theory on a circle has an exact global symmetry corresponding to KK momentum, since the dimensional reduction of the $(4,0)$ multiplet does not produce a KK photon, and is therefore likely in the Swampland. We also analyze other examples of exotic theories of gravity, and find that some of these (such as the $(3,1)$ theory in 6D and its non-supersymmetric cousins) evade our considerations, and may be realized as consistent quantum gravities. Since we were not able to exclude the $(3,1)$ theory in 6D, the question of string universality with 32 supercharges remains open, from the Swampland point of view\footnote{One can argue however that if the 5d theory has the duality symmetry of M-theory on $T^6$, the vector cannot couple to KK charge \cite{hull_unpub}. So if the possibility is realized it has to be in a disconnected component of moduli space, or perhaps in a non-supersymmetric version, as we discuss in Section \ref{sec:4}.}!  

More broadly, the goal of this note is to bring attention to the exotic gravities with spins $\leq 2$, including the non-supersymmetric ones. Some of them become ordinary Einsteinian theories upon dimensional reduction, and they seem to evade arguments about infinite towers in higher-spin theories \cite{Alba:2015upa}, so they could in principle have consistent interactions involving only a finite number of fields. We feel that some effort should be devoted to understanding their consistency and whether they appear somewhere in the bulk of the string Landscape. 

The paper is structured as follows. In Section \ref{sec:2}, we review the exotic $(4,0)$ theory and its circle compactification compactification to yield 5D $\mathcal{N}=4$ supergravity, as well as the recent literature on the topic. In Section \ref{sec:3}, we argue that supersymmetry together with exact Lorentz invariance force the theory to have an exact global symmetry, which puts it at odds with Swampland constraints. Section \ref{sec:4} generalizes the discussion to other exotic fields of spin $\leq 2$ in various dimensions. Finally, Section \ref{sec:5} contains a few concluding remarks.

\section{Review of the \texorpdfstring{$\mathcal{N}=(4,0)$}{N=(4,0)} theory and its relation to supergravity}\label{sec:2}
We start with a lighting review of the exotic $(4,0)$ supergravity. For our arguments, the basics about the field content and symmetry structure are sufficient; see \cite{Hull:2002hmj,deMedeiros:2002qpr,Minasian:2020vxn} for more details about dynamics and attempts to construct consistent interactions.

\subsection{Field content of 6D \texorpdfstring{$\mathcal{N}=(4,0)$}{N=(4,0)} and dimensional reduction}
    Maximally supersymmetric SUGRA in 6 spacetime dimensions has three possible realizations, with different chiralities in the SUSY algebra. The $\mN=(2,2)$ multiplet is the standard gravity multiplet, arising for example in the low energy regime of M-theory on $T^5$. In this note we will focus on the exotic $\mN=(3,1)$ and $\mN=(4,0)$ theories, which contain no graviton in their smallest multiplet.
    
    We here examine in detail the (4,0) multiplet, while we briefly comment on the (3,1) later on in Subsection \ref{sec:2.3}. The six-dimensional (4,0) superalgebra is given by \cite{Hull:2002hmj}
    \be \label{eq:alg6d}
        \begin{aligned}
            \{Q_\al^a,Q_\beta^b\} &= \Omega^{ab}(\Pi_+\Gamma^M C)_{\al \beta}P_M + (\Pi_+\Gamma^M C)_{\al \beta}Z^{ab}_M \\ &+ \frac{1}{6}(\Pi_+\Gamma^M\Gamma^N \Gamma^P C)_{\al \beta} Z^{ab}_{MNP}
        .\end{aligned}
    \ee
    Here $\al, \beta = 1,...,4$ are spinor indices, $M,N,P = 0,...,5$ are Lorentz indices and $a,b=0,...8$ are the internal $Sp(4)$ R-symmetry indices\footnote{In general, the R-symmetry group of the 6D $\mathcal{N}=(p,q)$ algebras is $\Sp(p)\times Sp(q)$.}. $\Omega_{ab}$ is the unique $Sp(4)$-invariant pairing, $C$ is charge conjugation, and $\Pi_{\pm}$ are the chiral projectors $\tfrac{1}{2}(1 \pm \Gamma_7)$.
    \\Lastly, $P_M$ is the six-dimensional momentum, and the central charges satisfy
    \be
        \begin{aligned}
            Z^{ab}_M = -Z^{ba}_M&, \quad \Omega_{ab}Z^{ab}_M=0 \\
            Z^{ab}_{MNP} = Z^{ba}_{MNP} = Z^{ab}_{[MNP]}&, \quad Z^{ab}_{MNP} = \frac{1}{3!}\epsilon_{MNPQRS}Z^{abQRS}.
        \end{aligned}
    \ee

Massless fields are then labeled by representations of the massless little group $SU(2)\times SU(2)$, together with the R-symmetry $Sp(4)$.  In terms of $SU(2)\times SU(2)\times Sp(4)$, the only multiplet containing fields of spins $\leq 2$ decomposes as \cite{Strathdee:1986jr}

    \be
        \textbf{(3,1;27)} \oplus \textbf{(1,1;42)} \oplus \textbf{(5,1;1)} \oplus \textbf{(4,1;8)} \oplus \textbf{(2,1;48)}.
    \ee
    The summands represent respectively:

    \begin{itemize}
        \item 27 self-dual 2-forms $B^i_{MN}$
        \item 42 scalars $\phi^i$
        \item an exotic rank 4 tensor $C_{MNPQ}$ with the symmetries described below
        \item 48 chiral fermions $\lambda^i$
        \item 8 fermionic self-dual 2-forms $\psi_{MN}$
    \end{itemize}
    The most immediate observation is that there is no graviton in the spectrum, already signaling that the theory is not gravitational in the conventional sense.
    \\The index structure of the tensor $C_{MNPQ}$ is that of the Riemann tensor: it has symmetries
    \be \label{eq:ctens}
        C_{MNPQ} = -C_{NMPQ} = -C_{MNQP} = C_{PQMN}, \hspace{1cm} C_{[MNP]Q}=0
    \ee
    It is sometimes convenient to view this as a symmetric tensor product of two 2-forms (later one will refer to such an object as a [2,2]-form) on which we impose the second condition above.
    In this language, the generalized notion of field strength is obtained by applying the exterior derivative on both 2-form components:
    \begin{align}
        G =& (\text{d}\otimes\text{d})C \\
        G_{MNPQRS} =& \partial_{ [M } C_{NP][QR,S]}.
    \end{align}
    The field strength satisfies the self-duality constraint:
    \begin{align} \label{eq:selfdual}
        G = (* \otimes 1)G = (1 \otimes *)G \\
        G_{MNPQRS} = \frac{1}{6}\epsilon_{MNPTUV}{G^{TUV}}_{QRS},
    \end{align}
    which we will soon see is crucial for dimensional reduction.
    There is also gauge symmetry that preserves $G$: one can shift $C$ by the the most general exact element with respect to the exterior derivative $(\text{d}\otimes\text{d})$ that preserves the properties \eqref{eq:ctens}. In components this is
    \be \label{eq:gauge}
        \delta C_{MNPQ} = \partial_{[M}\chi_{N]PQ} + \partial_{[P}\chi_{Q]MN} - 2 \partial_{[M}\chi_{NPQ]}
    \ee
    for a 3-tensor $\chi_{MNP}$ that is antisymmetric in the last two indices, $\chi_{MNP} = - \chi_{MPN}$.
    \\Since $C$ shares the algebraic properties of the Riemann tensor, it is  natural to ask if it is in general exactly the Riemann tensor for some ``implicit" metric $g^{(C)}$, which might be the actual, fundamental degree of freedom. This is not the case, even locally on a small patch\footnote{It can only be done at a single point, as can be shown e.g. using Riemann normal coordinates.}, as shown in \cite{deturck1986nonlinear}. 
    
    We now examine the reduction to 5 dimensions of the free theory, following \cite{Hull:2000rr,Hull:2002hmj}. We focus on the bosonic sector, as it already exhibits the features we are interested in. For the rest of the work, upper case indices $M,N...$ will describe fields in the higher dimensional theory, while greek indices $\mu, \nu...$ are used for extended dimensions in the lower dimensional theory.
    \\The 42 scalar fields survive in $D=5$. On a circle parametrized by $x_5$, each 2-form $B_{MN}$ reduces in the standard way to a vector $A_\mu = B_{\mu 5}$ and a 2-form $b_{\mu \nu} = B_{\mu \nu}$. If $B_{MN}$ is self-dual in $D=6$, then $A_\mu$ and $b_{\mu \nu}$ are dual to each other in $D=5$, so in lower dimension we can describe the theory only in terms of the vector field $A_\mu$, see e.g. \cite{Hull:2000rr} for an explicit derivation.
    \\More interesting is the behaviour of the exotic tensor $C_{MNPQ}$. It reduces to tensors of rank 2, 3 and 4 as

    \be\label{eq:dual}
        h_{\mu \nu} = C_{\mu 5 \nu 5} \hspace{1cm} d_{\mu \nu \rho} = C_{\mu \nu \rho 5} \hspace{1cm} c_{\mu \nu \rho \sigma} = C_{\mu \nu \rho \sigma}.
    \ee
    At the linearized level, gravity has a dual description \cite{Hull:1997kt,Hull:2000zn,Hull:2001iu} of the dynamics of the graviton $h_{\mu \nu}$ in terms of either a 3-form field $d_{\mu \nu \rho}$ or a 4-form field $c_{\mu \nu \rho \sigma}$, with precisely the same algebraic symmetries as the analogous fields in \eqref{eq:dual}. In fact it can be shown that the self-duality condition \eqref{eq:selfdual} imposes that these two pairs of tensor, obtained respectively by dimensional reduction of $C_{MNPQ}$ and by duality, coincide \cite{Hull:1997kt,Hull:2023iny}. Concretely, this means that the only physical degrees of freedom coming from $C_{MNPQ}$ in the compactified theory are those of a graviton. All in all, we found precisely the field content of 5-dimensional maximal SUGRA, given at the bosonic level by
    \begin{itemize}
        \item 27 abelian vector fields $A_\mu^i$ 
        \item 42 scalars $\phi^i$ parametrizing the homogeneous space $E_{6(6)}/Sp(4)$
        \item a graviton $g_{\mu \nu}.$
    \end{itemize}
    The fermions can also be shown to reduce to those of the 5D SUGRA, and the source-free equations of motion also match\cite{Hull:2000rr}.

    This match between the 5D and 6D theories is the main reason that led Hull \cite{Hull:2002hmj} to propose that the 6D (4,0) theory might arise as a strong coupling limit of 5D $\mN=4$ SUGRA. The mechanism should be analogous to how 5-dimensional $\mN=2$ Yang-Mills theory is described in the strong coupling regime as a 6D (2,0) superconformal field theory.

    Let us now discuss this matching at the level of the 5D and 6D superalgebras. The 5 dimensional $\mN=4$ superalgebra  reads
    \be \label{eq:alg5d}
        \begin{aligned}
            \{Q_\al^a,Q_\beta^b\} &= \Omega^{ab}(\Gamma^\mu C)_{\al \beta}P_\mu + \Omega^{ab}C_{\al \beta}K \\
            &+ (\Gamma^\mu C)_{\al \beta}Z^{ab}_\mu + C_{\al \beta}Z^{ab} +\frac{1}{2} (\Gamma^\mu \Gamma^\nu C)_{\al \beta} Z^{ab}_{\mu \nu}.
        \end{aligned}
    \ee
    There are now vector central charges $Z_\mu^{ab}$, 2-tensor central charges $Z_{\mu\nu}^{ab}$, and scalar central charges $K$ and $Z^{ab}$. These arise from reduction of the 6D charges and momentum, which decompose as \cite{Hull:2002hmj}
    
    \be
        \begin{aligned}
            P^M &\longmapsto (P^\mu,P^5 = K) \\
            Z_M^{ab} &\longmapsto (Z_\mu^{ab},Z_5^{ab}=Z^{ab}) \\
            Z^{ab}_{MNP} &\longmapsto Z_{\mu \nu 5}^{ab}=Z_{\mu \nu}^{ab} \\
        \end{aligned}.
    \ee
     The scalar charges $Z_{ab}$ are carried by electrically charged particles arising from 6D strings charged under $Z_\mu^{ab}$ and wrapped around the compact direction. The singlet charge $K$ is from the 6D point of view the Kaluza Klein charge associated to quantized momentum in the compact dimension; it will play a major role in our arguments in Section \ref{sec:3}.

\subsection{Adding interactions}
    One hurdle in making the 6D-5D connection described above more precise is that 5D SUGRA also has interactions, which are considerably harder to write down for the 6D (4,0) theory. It has in fact not yet been achieved, and there are multiple hints that if the interacting theory exists at all, it possesses very exotic properties. For instance, the $(4,0)$ analog of the S-matrix amplitudes constructed in \cite{Heydeman:2018dje} for the $(2,0)$ theory seems to have non-local behavior.\footnote{We thank M. Heydeman for explaining this point to us.} We give here a partial survey of the literature on this question.
    The 6D-5D connection discussed above has a field theory cousin in the well-known relationship between 5D $\mathcal{N}=2$ Yang-Mills theory, and its strong coupling limit as the 6D $\mathcal{N}=(2,0)$ superconformal field theory reduced on $S^1$. Even in this well-established case, writing down Lagrangian interactions is not possible (see, however, \cite{Henneaux:2017xsb,Henneaux:2018rub} for an action principle for the free theory).
    In \cite{Bekaert:1999dp,Bekaert:2000qx} the authors consider local, Lagrangian deformations of the free action of self-dual 2-form fields in 6 dimensions, that could give rise to the non-abelian interactions of Yang-Mills theory upon dimensional reduction. The classification of local Lagrangian deformations at first order in the coupling constant can be mapped exactly to a cohomology problem via the BRST formalism \cite{Barnich_1993}. This allowed the authors of \cite{Bekaert:2000qx} to rigorously show that local Lagrangian interactions in the 6D $\mN=(2,0)$ theory are necessarily abelian, and therefore the interactions in $D=6$ that generate the correct non-abelian ones in $D=5$ cannot come from a local Lagrangian. 
    Indeed, on M-theory grounds, the system is at strong coupling, and there is no good reason why interactions should be Lagrangian. The $\mathcal{N}=2$ SYM theory in 5D with gauge group $SU(n)$ is realized as the worldvolume theory of a stack of $n$ D4-branes in Type IIA string theory. At strong coupling this system is described as a stack of M5-branes wrapped on a circle of radius $R \sim g_{YM}$, whose worldvolume theory is in turn described by the superconformal $\mN = (2,0)$ field theory.
    
    The situation for the proposed decompactification limit of 5D $\mN=4$ SUGRA into the 6D $(4,0)$ theory is analogous to the Yang-Mills and $(2,0)$ in 6D at the level of the free theories, but it does not possess such a clear embedding into a string theory/M-theory framework. Local Lagrangian interactions in $D=6$ can be similarly shown to not suffice to generate the non-polynomial gravitational interactions in $D=5$ \cite{Bizdadea:2003ht,Bekaert_2005}, so the hypotetical full quantum $D=6$ theory should contain ``M-interactions" along the lines of the ones discussed above. 
    
 Reference \cite{Minasian:2020vxn} discusses a different question in the context of dimensional reduction. 
    In the 5D action there are cubic Chern-Simons terms \cite{Cremmer:1980gs} of the form
    \be \label{eq:cs}
        S_{CS} = \sum_{a,b,c}  \int \kappa_{abc} A^a \wedge F^b \wedge F^c
    \ee
    where the coefficients $\kappa_{abc}$ are constant and quantized, and the indices run over the \textbf{27} representation of $E_{6(6)}(\mathbb{Z})$ (the duality group of M-theory on $T^6$).  The question tackled by \cite{Minasian:2020vxn}  is that there is no obvious way to generate the Chern-Simons interactions of 5-dimensional $\mathcal{N}=4$ SUGRA from the 6D theory upon dimensional reduction.
    
    The authors of \cite{Minasian:2020vxn} consider all possible local, Lorentz invariant interactions among the 6D $\mathcal{N}=(4,0)$ fields that could give rise to \eqref{eq:cs} when the theory is put on a circle, and find no suitable ones. This of course leaves open the possibility of an interaction that either is non-Lagrangian or breaks Lorentz invariance, but such an interaction is hard to construct.
    \\Related to the problem of generating Chern-Simons terms is the fact that the $\mathcal{N}=(4,0)$ exotic tensor has a non-vanishing gravitational anomaly, as it is believed that only anomaly-free theories generate properly quantized Chern-Simons interactions upon dimensional reduction \cite{Corvilain:2017luj,Corvilain_2020}. The fermionic degrees of freedom can be similarly shown to be anomalous\cite{Lekeu_2021}. The anomaly itself is not necessarily fatal, as the theory is not gravitational and a priori needs not be diffeomorphism invariant. To put it another way, the gauge transformations of the 5D gauge fields in \eqref{eq:cs} uplift to exotic gauge transformations  \eqref{eq:gauge} of the tensor $C_{MNPQ}$, and not to 6D diffeomorphisms, which are \emph{not} gauged in the 6D $(4,0)$.   It was noted by Hull since the first proposal \cite{Hull:2002hmj,Hull:2022vlv} that a possibility is that the 6D theory does not possesses typical diffeomorphism invariance, but rather some form of generalization thereof. The symmetry would act on ``generalized coordinates" represented by 3-tensors $X^{MNP}$ at the linear level as $\delta X^{MNP} = \Xi^{MNP}$. This way in 5-dimensions one recovers ordinary diffeomorphisms $\delta x^\mu = \xi^\mu$ by
    
    \be
        x^\mu = X^{55\mu}\hspace{0.1cm}, \hspace{1.9cm} \xi^\mu = \Xi^{55\mu}
    \ee
    The correct mathematical description of such an object, similar to a manifold but locally described with 3-tensors instead of the usual coordinates, is unclear. However, the 6D gravitational anomaly may signal that it is troublesome to define the 6D theory on general manifolds different from flat space.

  Reference  \cite{Minasian:2020vxn} also proposed an interesting way to ``recycle'' the $(4,0)$ theory in the spirit of 12D SUGRA and F-theory, as an auxiliary description of a lower-dimensional system. The proposal, named ``h-theory", can be thought as as describing a $T^3$-fibered  6-dimensional manifold where the dynamics in the fibers is ``frozen". While it is an appealing possibility, we will not consider this interpretation in the rest of the paper, since it does not describe a six-dimensional theory of gravity.

    \subsection{The \texorpdfstring{$\mathcal{N}=(3,1)$}{N=(3,1)} multiplet}\label{sec:2.3}
    As mentioned at the beginning of this Section, the (4,0) is not the only exotic $\mN=4$ multiplet in 6 dimensions. There is also a $\mN=(3,1)$ chiral multiplet with similar features. In particular, it contains an exotic 3-tensor $D_{MNP}$ with the algebraic properties
    \be \label{eq:dtens}
        D_{MNP}=-D_{NMP}, \quad D_{[MNP]}=0.
    \ee
    The field strength is defined as
    \be
        S_{MNPQR} = \partial_{[M}D_{NP][Q,R]}
    \ee
    and satisfies a self-duality condition on its 3-form component,
    \be \label{eq:sd31}
        S_{MNPQR} = \frac{1}{6}\epsilon_{MNPTUV}{S^{TUV}}_{QR}.
    \ee
    This is similar to the (4,0) tensor in that it produces a graviton upon dimensional reduction, but with a crucial difference: there is a photon as well. Let us see this: naively one gets a vector, a 2-form, a graviton and a 3-tensor as
    \be
        a_\mu = D_{\mu 5 5}, \hspace{1cm} b_{\mu \nu}= D_{\mu \nu 5}, \hspace{1cm} h_{\mu \nu}=D_{\mu 5 \nu}, \hspace{1cm} d_{\mu \nu \rho} = D_{\mu \nu \rho}.
    \ee
    The self-duality constraint \eqref{eq:sd31} however identifies $b_{\mu \nu}$ with the dual of $a_\mu$ and $d_{\mu \nu \rho}$ with the dual of the graviton $h_{\mu \nu}$. The photon $a_\mu$ is an R-symmetry singlet, and could therefore couple with the central charge $K$ of \eqref{eq:alg5d}.
However, it seems that, if one assumes the 5d theory is the moduli space of M-theory on $T^6$, this photon can be shown  to couple to a particular linear combination of the 27 charges $Z_{ab}$ of \eqref{eq:alg5d} \cite{hull_unpub}, and therefore it would not be gauging KK momentum. Limits with BPS states carrying this charge have been classified \cite{Hull_1996} and they are decompactifications to either M-theory or Type IIB string theory, seemingly leaving no room for a decompactification to the (3,1) theory. The possibility still exists that there is a yet to be discovered component of 5d $\mathcal{N}=4$ supergravity moduli space, which does not decompactify to M-theory, and on which the $(3,1)$ proposal is realized. Alternatively, a non-supersymmetric theory such as those examined in Section \ref{sec:4} might instead realize the coupling of $a_\mu$ with the singlet charge $K$.
    We stress again that any version of the (3,1) theory still presents many of the same features of the (4,0) when it comes to defining the interacting theory: it is anomalous, no local Lagrangian interaction can give rise to 5D Chern-Simons terms \cite{Minasian:2020vxn}, and it is hard to conceive Lagrangian interactions that would lead to the non-polynomial 5D gravitational interactions.
    
\section{ A global symmetry in the \texorpdfstring{$\mathcal{N}=(4,0)$}{N=(4,0)} theory} \label{sec:3}

We have seen how the $(4,0)$ theory in six dimensions could conceivably lead to a consistent UV completion of maximal 5D supergravity, if certain additional assumptions (that the theory has no Lagrangian, and possibly not even a local description) hold. One very important feature of the 5D supergravity obtained after $S^1$ dimensional reduction is that there is no KK photon nor a radion field in the lower-dimensional supergravity. This puts it in stark contrast with the dimensional reduction of Einsteinian theories, where these fields come from dimensional reduction of the higher-dimensional metric. We will now argue that in particular the absence of a KK photon means that the five-dimensional theory possesses an exact $U(1)$ global symmetry, and is therefore incompatible with Swampland constraints\footnote{This is not the only global symmetry the free theory has; a global higher-form fermionic symmetry had already been identified in the free theory  \cite{Wang:2023iqt}, but it is expected to be broken by gravitational interactions.} \cite{vanBeest:2021lhn}.  
\\To argue for the presence of the global KK symmetry, our assumptions will be that: \begin{enumerate}
\item The 6D $(4,0)$ theory is Lorentz invariant (i.e. possesses a Lorentz-invariant S-matrix), and
\item It makes sense to consider the 6D theory on an $S^1$ background, a circle of radius $R$, and one can construct states of this background by taking flat space 6D states and impose invariance under a $2\pi R$ translation.
\end{enumerate}
Notice that both assumptions are automatic in QFT and Einsteinian theory, and even in non-Einsteinian models such as little string theories \cite{Aharony:1999ks} or Vasiliev-type theories of gravity. Nevertheless, since we do not have a concrete model for the $(4,0)$ theory, we list them here as explicit assumptions.

Let us consider the theory on $\R^6$. As a consequence of the first assumption, the Hilbert space of the theory decomposes into representations of the Poincar\'{e} group, labeled as usual by a mass $m$ and spins. In six dimensions, the massless little group is $SU(2)\times SU(2)$, so states of massless particles are labelled by two spins $(j_1,j_2)$, and a momentum $p$, satisfying
\begin{equation}p^2=0.\end{equation}
This is the case for all the states in the free $(4,0)$ supermultiplet. Introducing interactions cannot make these states massive, since they live in a short multiplet. Notice that we made no assumption about the nature of the interactions or the Hilbert space itself. Our only assumption is that the $(4,0)$ supersymmetry acts nontrivially on states.

Now consider the theory on $\mathbb{R}^5\times S^1$, where the $S^1$ factor is a circle of radius $R$. Quantizing a theory on  $\mathbb{R}^5\times S^1$ is in general qualitatively different from quantizing on $\mathbb{R}^6$ (think of winding states in string theory). Furthermore, if the interactions in the $(4,0)$ are superconformal as proposed in \cite{Hull:2002hmj}, the size $R$ is the only dimensionful parameter in the theory, and must equal the 5D Planck length\footnote{It is difficult to see how exact conformality can be reconciled with the expected properties of black hole entropy and of the inelastic cross-section at very high center of mass energies and fixed impact parameter, where it should be dominated by black holes. However these considerations are independent of the arguments given in the main text.}. In any case, however, it is possible to regard the $S^1$ as the quotient of a real line parametrized by coordinate $x_5$ by translations
\begin{equation}x_5\sim x_5+2\pi R.\label{empy0}\end{equation}
Accordingly, by our second assumption, one can obtain valid quantum states of the theory on $\mathbb{R}^5\times S^1$ by taking $\mathbb{R}^6$ states invariant under \eq{empy0}. In particular,  a momentum eigenstate $\ket{p_\mu}$ is not invariant, since under translation by $2\pi R$ we get
\begin{equation} T_{2\pi R} \ket{p_\mu}= e^{2\pi i R p_5} \ket{p_\mu},\end{equation}
where $p_5$ is the component of momentum along the $S^1$ direction. This state is invariant provided that the momentum satisfies 
\begin{equation} p_5=\frac{n}{R}\end{equation}
for some integer $n$. From the five dimensional point of view, we obtain a tower of states of masses $m_n=n/R$ -- we have obtained a Kaluza-Klein tower of states for each 6D massless one. 
Notice that the states constructed in this way do not necessarily form a complete set; for instance, winding states in string theory constitute a counterexample. The point is that KK states like \eqref{empy0} will always be present, irrespectively of the form of the 6D interactions, as long as Lorentz symmetry is preserved.

The basic consequence of the above assumptions is that each $(4,0)$ 6D field results in a tower of 5D KK states, labelled by an integer KK momentum. All this is what one gets from the usual field theory discussion, but the argument above emphasizes that it relies on symmetries only, and therefore it applies to any avatar of the $(4,0)$ theory (local, nonlocal, etc.) in which there is a notion of physical Hilbert space carrying a representation of the superalgebra. 

Furthermore, from the point of view of the 5-dimensional theory, KK momentum is an exactly conserved charge. It cannot be spontaneously broken (meaning that the vacuum is not invariant), because it is a central charge for 5d $\mathcal{N}=4$ algebra, as we saw in Section \ref{sec:2}, and we assumed that the vacuum is maximally supersymmetric and thus annihilated by all supercharges. If it is an exact symmetry, Swampland constraints then demand that there is a massless $U(1)$ gauge field coupled to it. But there is no suitable candidate among the fields in the 5D $\mathcal{N}=4$ multiplet. As studied in Section \ref{sec:2}, the only massless vectors in five dimensions come from the dimensional reduction of the 6D $(4,0)$ two-forms. The electric couplings $\int A$ come from dimensional reduction of $\int B$, where $B$ are the 6D self-dual 2-forms. Therefore charged particles can only come from self-dual strings in the 6D theory, and therefore the gauge fields are not gauging KK momentum. The 6D strings are the same that provide the magnetically charged objects under the 27 fields in the five-dimensional theory, so they are necessarily present in the theory via the Completeness Principle \cite{Polchinski:2003bq}. Additionally, if we if one takes the $E_{6(6)}(\mathbb{Z})$ duality symmetry of M-theory on $T^6$ exactly, none of the 2-forms can become the KK photon, since the latter is a singlet under duality and the 2-forms transform in the $\mathbf{27}$.

 We reach the conclusion that there is an exactly conserved KK charge, but no gauge boson coupled to it. This is a global symmetry, and therefore, the minimal interacting $(4,0)$ theory is in the Swampland. As a first check of this claim, the actions constructed in \cite{Henneaux:2017xsb,Henneaux:2018rub} for the free (4,0) and (3,1) theories are invariant under this symmetry, since they preserve supersymmetry, and thus also translations in the compact direction.

To evade this conclusion, one must negate one of the two assumptions above; either the theory does not carry a representation of the $(4,0)$ algebra in its Hilbert space (but then, in which sense is this a $(4,0)$ theory?), or the flat space and $S^1$ compactifications cannot be related to each other in the simple way above, so there are no KK charged states and the KK charge operator is the identity. However, we remark again that this construction (the ability to obtain some states of the theory quantized in $S^1$ by looking at translation-invariant states of the theory on $\mathbb{R}$) holds true in QFT, string theory, and more exotic constructions. Negating it is tantamount to declaring that the $S^1$ compactification of the $(4,0)$ theory does not resemble the flat space theory in any way. At that point, we might as well just think of them as two completely separate theories that only have in common the number of supercharges.

It is also interesting to ask what happens when we compactify more than one dimension\footnote{We thank Chris Hull for bringing up this question.}, for instance, when the $(4,0)$ is compactified on a square $T^2$. In the four-dimensional theory, there is only one KK photon that can couple to KK charge, but there are two towers of KK modes; so one of them is still global. The $\mathbb{Z}_4$ rotational symmetry of the square torus acts on KK states with no invariant subspace, so it must be broken (either spontaneously or explicitly) by the interactions. This is perhaps not surprising given that the 6d theory has a diffeomorphism anomaly \cite{Minasian:2020vxn}. However, unlike an ordinary anomaly, the breaking effects naively disappear in the free theory; it seems the symmetry is explicitly broken merely by the assumption that Einsteinian interactions appear when the $(4,0)$ theory is compactified. 

Reference \cite{Hull:2022vlv} discusses a proposal for the supergravity current coupled to the charge $K$.
 By analogy to the field theory case (where maximal 5D SYM is the low-energy limit of the $(2,0)$ theory on $S^1$, and the solitonic Yang-Mills instanton charge $\text{tr}(F^2)$ actually corresponds to a smeared version of KK momentum), it is proposed that the current
\begin{equation} j=*\text{tr}(R^2), \label{w23}\end{equation}
which is automatically conserved in supergravity, may be a smeared version of the KK charge $K$. This identification is then supported by the fact that there are singular ALE geometries with charge \eqref{w23}, which, in analogy with the Yang-Mills case, may be identified with the KK modes themselves.

\begin{figure}[!htb]
\begin{center}
\includegraphics[width=8.5cm]{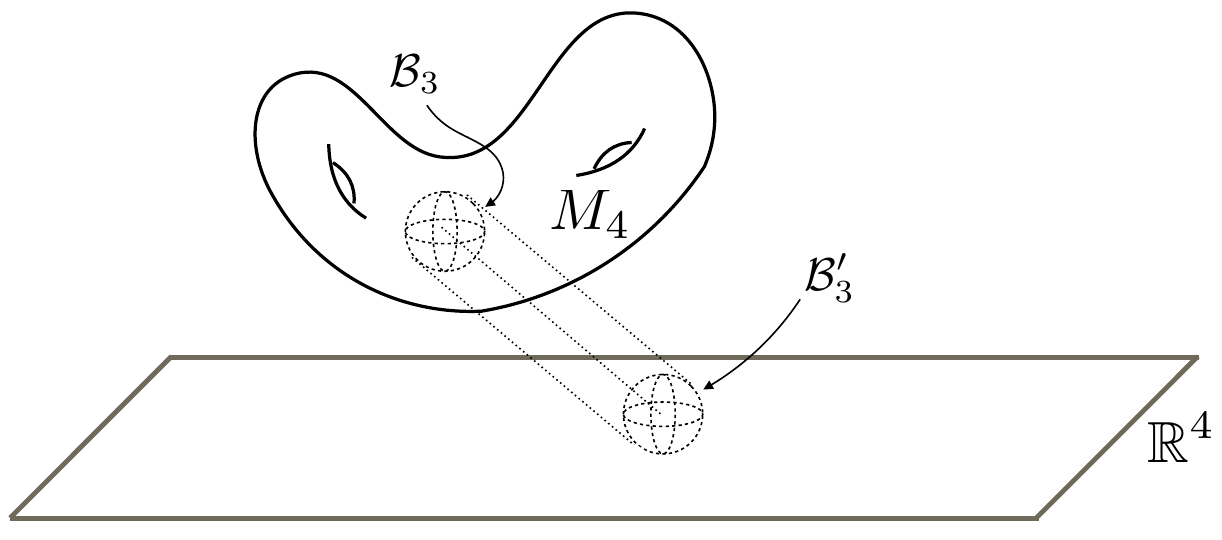}
\end{center}
\caption{To construct an asymptotically flat particle-like soliton with nonzero $\text{tr}(R^2)$, take an arbitrary 4-manifold, remove a little 3-ball $\mathcal{B}_3$ at an arbitrary point, then glue it to $\mathbb{R}^4$ minus a similar little ball $\mathcal{B}'_3$ at the origin.}
\label{f1}
\end{figure}

We will now argue that the current \eqref{w23} is broken, and therefore cannot couple to $K$. More generally, one could ask whether \eq{w23} is a conserved charge in any supergravity of any dimension, and indeed, this question has been studied recently in a number of stringy examples, in connection with the Cobordism Conjecture \cite{McNamara:2019rup}. It turns out that the conservation of \eq{w23} can be violated by appropriate singular configurations. As explained in \cite{McNamara:2019rup}, one can construct examples of asymptotically flat manifolds with charge \eq{w23} by taking a compact 4-manifold with nonzero $\int \text{tr}(R^2)$ (for instance, K3) and taking a connected sum with flat space (see Figure \ref{f1}). Asymptotically flat configurations charged under \eq{w23} are necessary if this current to be identified with a ``fattened out'' version of KK modes, since KK modes exist in flat space. The manifold thus constructed cannot be smoothly deformed to empty space while preserving the asymptotics, since that would change the value of $\int \text{tr}(R^2)$, which is quantized on a compact manifold (the gluing to $\mathbb{R}^4$ is irrelevant in this case since we assume all fields fall off sufficiently rapidly at infinity). In fact, $\int \text{tr}(R^2)$ would be preserved by even more complicated topology-changing processes  described via cobordisms (which are generically expected in quantum gravity), such as the ones depicted in Figure \ref{f1}, since it is a cobordism invariant. 

\begin{figure}[!htb]
\begin{center}
\includegraphics[width=6cm]{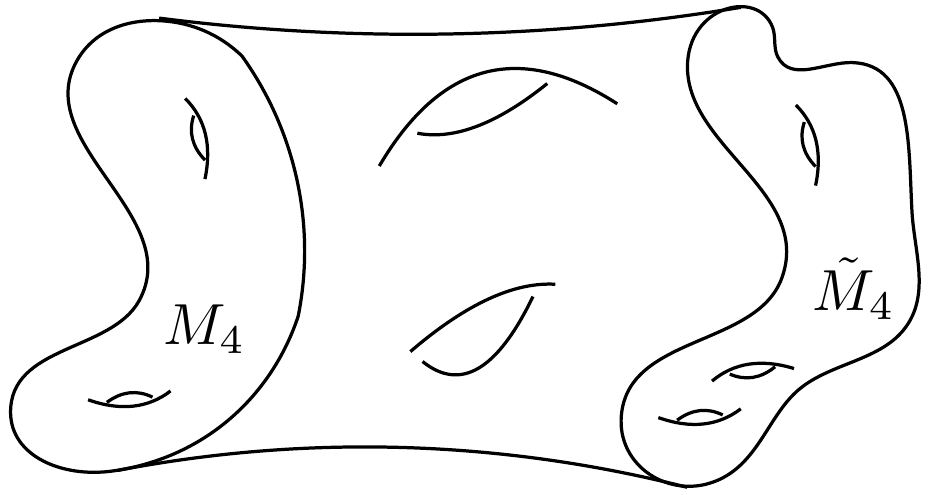}
\end{center}
\caption{Topology-changing processes may alter the topology of any particular soliton like the one constructed in Figure \ref{f1}, but as long as they can be described via smooth manifolds,  $\int \text{tr}(R^2)$ will be conserved, since it is a cobordism invariant.}
\label{f2}
\end{figure}

However, if we are doing this in a 5D $\mathcal{N}=4$ SUGRA that arises from the $(4,0)$ theory on $S^1$, then secretly our K3 is more appropriately described by $K3\times S^1$, where $S^1$ is the compactified additional circle. It turns out that, since the bordism group
\begin{equation}\Omega_5^{\text{Spin}}=0,\end{equation}
so there are cobordisms that can mediate a transition between $K3\times S^1$ and empty space. In other words, the topological charge is not conserved. From the five-dimensional point of view, the geometries that do this are singular (the $S^1$ goes to zero size at some point), but the point is that the $\text{tr}(R^2)$ charge is no longer conserved. The cobordism conjecture requires that singular configurations that can get rid of $\text{tr}(R^2)$ charge exist in any quantum theory of gravity (which bypasses the tricky question of whether one should indeed sum over topologies in the 6D $(4,0)$ theory). As a final comment in this regard, there is a strong parallel between 5D maximal SUGRA and $(4,0)$ on one hand and IIA strings and M-theory on the other, since both involve a $S^1$ reduction. The $\text{tr}(R^2)$ charge is indeed not conserved in IIA, precisely because of the effects mentioned above.

A related puzzle, which also arises in both 5D maximal SUGRA and IIA strings, is that circle reduction requires objects where $\text{tr}(R^2)$ measures an exactly conserved charge: Any Einstein theory of gravity reduced on $S^1$ includes a KK photon and magnetic monopoles for it, described by Taub-NUT spacetimes whose monopole ``nut charge'' is measured precisely by a nonzero $\text{tr}(R^2)$. How is this compatible with the claim above that $\text{tr}(R^2)$ is broken? The answer is that the argument in the previous paragraph crucially relies on flat Euclidean asymptotics, and it only shows non-conservation of $\text{tr}(R^2)$ in this case. It does not apply in cases where the asymptotic geometry is not $\mathbb{R}^4$. In particular, it is not possible to compactify the sphere times circle at infinity in a Taub-NUT space in a trivial way (i.e. as a 3-disk times $S^1$), due to the fact that, in a Taub-NUT background, the $S^1$ is non-trivially fibered over the $S^2$ at infinity \cite{Witten:2009xu}. In this way, it is possible to have  $\text{tr}(R^2)$ be an exactly conserved charge in some situations and not in others. Unfortunately, the one required for agreement with the $(4,0)$ uplift is the one where it is broken. 

We end up this Section on a positive note. By contrast to the $(4,0)$ theory, the $(3,1)$ exotic multiplet, which we described in Section \ref{sec:2.3}, does produce a vector when reduced on a circle, coming directly from reduction of the exotic tensor in the multiplet.

We cannot exclude that this vector couples to KK charge, as the KK photon does in ordinary gravity theories (however, see footnote in the Introduction). If this is the case, the corresponding symmetry would be gauged, and the $(3,1)$ theory is not excluded by the no global symmetries conjecture. Of course, as we already remarked, there are other considerations, such as the existence of consistent interactions, that may pose additional challenges. But it is interesting that everything seems fine from the Swampland point of view.

\section{Reduction of other exotic tensors}\label{sec:4}
    In the previous Section, we saw that the  5D Einsteinian theory arising from the 6D (4,0) on a circle has a global symmetry that relegates it to the Swampland. This feature only relies on the index structure of the exotic 4-tensor $C_{MNPQ}$ and the self-duality conditions that remove the extra degrees of freedom.
    We could therefore try to identify other similar tensors that reduce to a gravitational theory upon compactification, independently of whether they appear in a supersymmetric multiplet or not. 
    In this Section, we consider free theories of a single exotic tensor in generic dimension $D$ and their reduction on tori of arbitrary dimension. For a torus of sufficiently high dimension, there will only be ordinary fields of spins $\leq 2$. After the reduction, we will demand that
    \begin{itemize}
        \item There is not more than one graviton,
        \item There are $d$ vector fields, in the case of reduction on $T^d$, that could in principle play the role of the required KK photons.
    \end{itemize}
    We demand the first point since we do not know how to construct consistent interactions with more than one graviton \cite{Boulanger:2000rq}. The second comes from the generalization of our argument in the previous section.

    The language of Young tableaux can be used to describe representations of the orthogonal group, and it also turns out to be convenient to describe both the phenomenon of self-duality and that of dimensional reduction, so we briefly review it here (see also \cite{Hull:2001iu,BURD_K_2001}).
    A Young tableaux is a collection of $m$ boxes arranged in rows and columns of decreasing length. For our purposes, we associate a tableaux to a representation of $SO(D-2)$, the little group of massless fields in $D$ dimensions. Each box represents a factor isomorphic to the fundamental representation $V$ of the little group $SO(D-2)$ in the tensor product representation $V^{\otimes m}$, or in other words, a Lorentz index. The indices associated to factors in the same column are antisymmetrized, and indices along the same row are symmetrized.
    Representations (anti)symmetrized in this way are in general not irreducible yet, and some ``tracelessness" conditions have to be applied, as in the familiar case of the graviton\footnote{$SU(N)$ and $Sp(N)$ also have representations described by Young diagrams. In the case of $SU(N)$ the symmetry structure is enough to guarantee irreducibility; $Sp(N)$ representations, analogously to $SO(N)$, require additional conditions, see e.g. part III of \cite{fulton1991representation}.}. These do not change the index symmetry structure of the representation, and each such structure is associated to a single irreducible representation.
    For example, the tableaux associated to scalar, vector, graviton and $p$-form representations are respectively
    
    \begin{equation}
        \ytableausetup{smalltableaux}
        \textbf{1},\quad \ydiagram{1},\quad
        \ydiagram{2} , \quad
        \begin{ytableau}
            \ \\
            \ \\
            \none[:] \\
            \
        \end{ytableau}  \,,
    \end{equation}
    where the column has $p$ rows.
    
    Dimensional reduction $D\rightarrow D-1$ is simple to visualize in this language: indices in the compact directions do not transform under the lower dimensional little group $SO(D-1-2)$, so for each such index we remove the corresponding box. Each individual index may be put in the compact dimension or not, so one needs to consider both cases for each index. However, because of antisymmetry, we cannot put two indices from the same column in the compact direction, and we are therefore allowed to remove at most one box from each column. Symmetry along the rows might identify some of these removals.
    As an example, let us represent the usual reduction of a graviton along a circle to produce a graviton, the KK photon, and the radion:
    \begin{equation}\label{yt:gravred}
        \begin{ytableau}
            \ & \ \\
        \end{ytableau}
        \hspace{1mm}
        \longmapsto
        \hspace{1mm}
        \begin{ytableau}
            \ & \ \\
        \end{ytableau}
        \hspace{1mm}\scalebox{1.}{$\bigoplus$} \hspace{1mm}
        \begin{ytableau}
            \
        \end{ytableau}
        \hspace{1mm}  \scalebox{1.}{$\bigoplus$} \hspace{1mm}
        \textbf{1}.
    \end{equation}
    Another ingredient that has to be taken into account is duality. $SO(D-2)$ has two invariant tensors: a symmetric 2-tensor $\delta$ (the metric), which we used to take traces above, and a totally antisymmetric rank $D-2$ tensor $\epsilon$, which can be used to identify different representations with different Young tableaux. At the level of differential forms, this is the familiar Hodge duality between $p$-forms and $(D-2-p)$-forms in $D$ dimensions. Since the Young tableaux for a $p$-form is a single column of length $p$, the dualization procedure simply corresponds to swapping a column of length $p$ with one of length $D-2-p$. For general tensors, this duality can be applied to each totally antisymmetric component, associated to a column in the Young diagram.  When $D-2$ is even, duality is an endomorphism on the space of $(D-2)/2$-forms. When $D-2$ is a multiple of 4, this endomorphism squares to $+1$, so the $(D-2)/2$-form representation becomes reducible, and can be decomposed into self-dual and anti-self-dual parts. This self-duality constraint can be imposed separately in every column of length $(D-2)/2$ of a general Young tableaux, although due to symmetry among columns the same choice between self-dual and anti-self-dual should be made in all of them.
    
    When reducing on a circle, a self-duality constraint in the higher dimension can ``trickle down" and identify among each other some representations which naively looked independent in the dimensional reduction decomposition. Let us show this in the main example of this work, the exotic tensor $C_{MNPQ}$. This tensor is represented by the Young tableaux
    \begin{equation}
        \begin{ytableau}
            \ & \ \\
            \ & \ \\
        \end{ytableau}
    \end{equation}
    on which a self-duality constraint is imposed on both columns. Ignoring this constraint, the naive dimensional reduction would look like
    
     \begin{equation}\label{eq:4.3}
        \begin{ytableau}
            \ & \ \\
            \ & \ \\
        \end{ytableau}
        \longmapsto
        \begin{ytableau}
            \ & \ \\
            \ & \ \\
        \end{ytableau}
        \hspace{1mm} \scalebox{1.5}{$\oplus$}\hspace{1mm}
        \begin{ytableau}
            \ & \ \\
            \ \\
        \end{ytableau}
        \hspace{1mm} \scalebox{1.5}{$\oplus$}\hspace{1mm}
        \begin{ytableau}
            \ & \ \\
        \end{ytableau}
        .
    \end{equation}
     Upon imposing the $D=6$ self-duality constraint \eqref{eq:selfdual}, the three representations on the right hand side are actually all identified to give a single graviton, as discussed under \eqref{eq:dual}.
    \\As a last warmup before discussed the general case, let us consider the tensor $D_{MNP}$ that appears in the six-dimensional $\mN=(3,1)$ theory: it reduces as
    \begin{equation}\label{eq:4.4}
        \begin{ytableau}
            \ & \ \\
            \ \\
        \end{ytableau}
        \longmapsto
        \begin{ytableau}
            \ & \ \\
            \ \\
        \end{ytableau}
        \hspace{1mm} \scalebox{1.5}{$\oplus$}\hspace{1mm}
        \begin{ytableau}
            \ & \ \\
        \end{ytableau}
        \hspace{1mm} \scalebox{1.5}{$\oplus$}\hspace{1mm}
        \begin{ytableau}
            \ \\
            \ \\
        \end{ytableau}
        \hspace{1mm} \scalebox{1.5}{$\oplus$}\hspace{1mm}
        \begin{ytableau}
            \ \\
        \end{ytableau} 
        \xrightarrow[\text{duality}]{\text{self}}
        \begin{ytableau}
            \ & \ \\
        \end{ytableau}
        \hspace{1mm}        \hspace{1mm} \scalebox{1.5}{$\oplus$}\hspace{1mm}\hspace{1mm}
        \begin{ytableau}
            \ \\
        \end{ytableau}
    \end{equation}
    to give a graviton and a photon in the 5 dimensional theory. This theory evades the argument of Section \ref{sec:3}, since the photon could in principle couple to the KK charge, to gauge the corresponding symmetry. We however remark again that the formulation of a potential interacting theory presents other issues, as discussed in Sections \ref{sec:2.3}.
    
We have now seen how dimensional reduction works, but how do we restrict to Einsteinian theories, which only propagate fields of spin $\leq 2$? For massless particles, physical spin (or helicity) is defined by the eigenvalue that the state picks upon rotation around a plane transverse to the momentum -- and this is precisely captured by the little group representations that we have been discussing. For instance, in four dimensions, there is a single transverse plane, and correspondingly $SO(D-2)=SO(2)=U(1)$. The physical spin is the $U(1)$ charge of the little group representation, and for a representation with $n$ rows
\begin{equation}
    \underbrace{\begin{ytableau}
            \ & \ & \none[] & \none[ ...] & \none[] & \ \\
    \end{ytableau}}_{n}
\end{equation}
the spin is equal to the number of rows (in this case, there are no columns, since there is a two-index antisymmetric tensor and only two free indices available). In higher dimensions, there is more than one transverse plane available, and a given physical field can have polarizations with different physical spins. These can be found simply by decomposing the $SO(D-2)$ representation into $SO(2)$ representations corresponding to the 2-plane of interest. Since dimensional reduction can reduce, but never increase the number of rows, the maximum helicity propagated by any field in any number of dimensions equals the number of rows in its Young tableaux. 
    
The upshot of this discussion is that, if we are interested in theories that become Einsteinian when reduced to five dimensions or less,  we can limit ourselves to fields whose Young diagram has at most two columns, of possibly different lengths $k$ and $l$, where without loss of generality we take $k \geq l$. In other words, tableaux of the form 
    \begin{equation}\label{yt:3}
        \ytableausetup{smalltableaux}
      \begin{ytableau}
            \ & \ \\
            \ & \ \\
            \none[:] & \none [:]\\
            \ & \ \\
            \none[:]\\
            \
        \end{ytableau}
    \end{equation}
with first column of length $k$, second of length $l$. As we will now see, the requirement of having only one graviton in the low-dimensional spectrum is enough to exclude all representations of this form except for those we already examined, namely the 4-tensor $C_{MNPQ}$ that appears in the (4,0) theory and the 3-tensor $D_{MNP}$ of the (3,1) theory.
    The key observation is that self-duality is necessary to have a single graviton. Let us see what happens if one does not impose it. Consider the case of a $[k,l]$ diagram in $D \geq 2k+2$. This is completely general, since if any column if longer than $(D-2)/2$, one can dualize to a shorter one. As can be seen in the example of \eqref{eq:4.4}, without imposing self-duality a $[k,l]$-type diagram reduces to at least two distinct two-column representations, namely a $[k,l]$ and a $[k-1,l]$ (if $l>1$ there will be generically be even more, but two are sufficient for our argument).
    If we continue compactifying down to four dimensions, each of these two column objects will eventually reduce to a graviton, obtained by taking boxes off both columns until they reach length 1 and dualizing at any step if possible to get a shorter column.
    
    Imposing a self-duality condition on the original representation can avoid the presence of more than one two-column object by identifying them after compactifying on $S^1$. Assuming this happens (we show in Appendix \ref{sec:A} that in general it does not), the resulting $[k-1,l]$ representation is however not self-dual anymore (simply because the compactified theory lives in odd dimension $D=2k+1$, where no self-duality condition exists). Therefore any further compactification presents the issue just discussed and will give at least two gravitons.

    The only way to avoid this is by imposing that the unique two-column representation obtained after the first circle compactification is the graviton already, such that any further reduction will look like the ordinary \eqref{yt:gravred}. This needs the length $k-1$ of the longest column after compactification to be 1, which in turn fixes
    \begin{equation}
            k = 2, \quad
            D = 2k+2 = 6.
    \end{equation}
    The only two options are then the exotic tensors in the 6-dimensional (4,0) and (3,1) theories. Any other tensor necessarily will produce more than one graviton in 5 dimensions or lower.
   
One may also wonder which theories comply with the absence of global symmetries alone, without imposing a bound on the number of gravitons. In Appendix \ref{sec:A} we perform this general analysis, and show that ``L-shaped" diagrams of type $[k,1]$, when compactified on a torus of dimension $k-1$, give rise to exactly $k-1$ photons, the correct number to couple to the KK charges. They generalize the exotic tensor of the 6D $(3,1)$ multiplet in that they evade the global symmetry argument of Section \ref{sec:3}. The presence of exotic tensor fields however leads to multiple gravitons upon further compactifications, and so they do not seem to be of physical interest.

\section{ Similar proposals in the Swampland and final comments}\label{sec:5}

Swampland Conjectures are most often applied to models of Einsteinian gravity (see \cite{Klaewer:2018yxi,Heisenberg:2019qxz,Basile:2021krr,Eichhorn:2021qet} for some exceptions). The (4,0) theory is certainly not Einsteinian, and in fact, it does not even contain a metric among its degrees of freedom. However, since it reduces to an ordinary theory of gravity when put on a circle, it is affected by Swampland constraints. In this note we have explained how the theory contains an exact KK global symmetry, due to the absence of a KK photon, and hence, the theory is in the Swampland. 

One may ask about ways to evade this conclusion. Some possibilites are simply forbidding the $S^1$ compactification or assuming that the KK towers of 6D states are absent for some reason. But other than this, the argument is pretty general. In a sense, the $(4,0)$ proposal, assuming suitable interactions existed in 6D, is an example of the asymptotic safety scenario, where a non-renormalizable gravity theory arises from a conformal fixed point and the Planck mass is dynamically generated. Here we see that, at least this particular instance of asymptotic safety is in the Swampland. This is to be contrasted with the ``field theory asymptotic safety'' scenario of 5D maximal SYM being UV-completed by the $(2,0)$ SCFT on a circle, which is actually realized in string theory. 

Other similar scenarios are not so easily discarded. In 6 dimensions, the exotic field of the $(3,1)$ supermultiplet produces a graviton and a KK-like photon upon dimensional reduction, so it avoids the simple argument that applies to the $(4,0)$ theory. Therefore, as far as we can tell, the theory might be realized at high energies in some particular locus of moduli space, if the obstacles to constructing consistent interactions can be overcome.

More generally, apart from supersymmetry, there is a very interesting question of whether perturbative theories of a finite number of exotic fields of spins $\leq 2$ might exist in various dimensions. Weakly coupled fields of higher spins can be argued on very general grounds to appear only in infinite towers, like a perturbative string (see e.g. \cite{Maldacena:2011jn,Camanho:2014apa}). Interestingly, this conclusion does not obviously hold for exotic fields of spin $\leq 2$ \cite{Alba:2015upa}. These exotic fields are an exciting arena for further exploration. For instance, one could look at the AdS analog of the theories described here, and study the putative CFT dual. This would be a conformal theory without stress-energy tensor, but with a conserved quantity that would become a $T_{ab}$ under dimensional reduction! At worst, if they are all inconsistent, showing this inconsistency explicitly may lead to new Swampland principles or techniques. At best, we may uncover a completely new phase of gravity! 

\section*{Acknowledgements}
We are very thankful to Jose Calder\'{o}n-Infante, Matt Heydeman, Chris Hull, Ruben Minasian and Alexander Zhiboedov, for enlightening discussions. MM thanks the IPhT Saclay workshop ```Deconstructing the String Landscape -- Landscapia'' for hospitality and providing a stimulating environment for discussion, as well as the KITP Program ``What is String Theory?''. This research was supported in part by grant NSF PHY-2309135 to the Kavli Institute for Theoretical Physics (KITP). MM is supported by an Atraccion del Talento Fellowship 2022-T1/TIC-23956 from Comunidad de Madrid. The authors also gratefully acknowledge the support of the grants CEX2020-001007-S and PID2021-123017NBI00, funded by MCIN/AEI/10.13039/501100011033 and by ERDF A way of making Europe. MT is supported by the FPI grant  CEX2020-001007-S-20-3 from Spanish National Research Agency from the Ministry of Science and Innovation.

\appendix
\section*{Appendix}
\section{Reduction of general exotic spin 2 fields}\label{sec:A}
Here we study the number of gravitons and KK photons produced from an exotic spin 2 tensor after dimensional reduction in full generality, without demanding that there is a single graviton on the spectrum in 4D. We consider separately the three types of diagrams represented in \eqref{yt:4}.
\begin{equation}\label{yt:4}
        \ytableausetup{smalltableaux}
        \begin{ytableau}
            \ & \ \\
            \ & \ \\
            \none[:] & \none [:]\\
            \ & \
        \end{ytableau}
        \hspace{0.5cm},\hspace{0.5cm}
        \begin{ytableau}
            \ & \ \\
            \ & \ \\
            \none[:] \\
            \
        \end{ytableau} 
        \hspace{0.5cm},\hspace{0.5cm}
        \begin{ytableau}
            \ & \ \\
            \ \\
            \none[:] \\
            \
        \end{ytableau} 
    \end{equation}
In Section \ref{sec:4} we already discussed how when reducing to $D=4$ we find more than one graviton. By reducing a $[k,l]$-type field from dimension $D=2k+2$ to $D=k+3$ we can however find theories with a single graviton and different amounts of photons, allowing to identify a global KK symmetry as in Section \ref{sec:3}.

Let us start from the last category in \eqref{yt:4}, as the dimensional reduction of ``L-shaped" diagrams is somewhat more straightforward. It is instructive to look at the first non-trivial example, namely $k=4$: we start from an 8-dimensional theory, with little group $SO(8)$, and reduce on a circle.
    
    \begin{equation}\label{eq:4.7}
        \begin{ytableau}
            \ & \ \\
            \ \\
            \ \\
            \ \\
        \end{ytableau}
        \longmapsto
        \begin{ytableau}
            \ & \ \\
            \ \\
            \ \\
            \ \\
        \end{ytableau}
        \hspace{1mm}\scalebox{1.5}{$\oplus$}\hspace{1mm}
        \begin{ytableau}
            \ & \ \\
            \ \\
            \ \\
        \end{ytableau}
        \hspace{1mm}\scalebox{1.5}{$\oplus$}\hspace{1mm}
        \begin{ytableau}
            \ \\
            \ \\
            \ \\
            \ \\
        \end{ytableau}
        \hspace{1mm}\scalebox{1.5}{$\oplus$}\hspace{1mm}
        \begin{ytableau}
            \ \\
            \ \\
            \ \\
        \end{ytableau} 
        \xrightarrow[\text{duality}]{\text{self}}
        \begin{ytableau}
            \ & \ \\
            \ \\
            \ \\
        \end{ytableau}
        \hspace{1mm}\scalebox{1.5}{$\oplus$}\hspace{1mm}
        \begin{ytableau}
            \ \\
            \ \\
            \ \\
        \end{ytableau}
    \end{equation}
    We have then no more duality to play with, and the reduction on an additional circle to $D=8$ is the naive one:
    
    \begin{equation}\label{eq:4.8}
        \begin{aligned}
            \begin{ytableau}
                \ & \ \\
                \ \\
                \ \\
            \end{ytableau}
            &\longmapsto
            \begin{ytableau}
                \ & \ \\
                \ \\
                \ \\
            \end{ytableau}
            \hspace{1mm}\scalebox{1.5}{$\oplus$}\hspace{1mm}
            \begin{ytableau}
                \ & \ \\
                \ \\
            \end{ytableau}
            \hspace{1mm}\scalebox{1.5}{$\oplus$}\hspace{1mm}
            \begin{ytableau}
                \ \\
                \ \\
                \ \\
            \end{ytableau}
            \hspace{1mm}\scalebox{1.5}{$\oplus$}\hspace{1mm}
            \begin{ytableau}
                \ \\
                \ \\
            \end{ytableau}
            \\
            \begin{ytableau}
                \ \\
                \ \\
                \ \\
            \end{ytableau}
            &\longmapsto
            \begin{ytableau}
                \ \\
                \ \\
                \ \\
            \end{ytableau}
            \hspace{1mm}\scalebox{1.5}{$\oplus$}\hspace{1mm}
            \begin{ytableau}
                \ \\
                \ \\
            \end{ytableau}
        \end{aligned}
    \end{equation}
    Further reduction look like copies of \eqref{eq:4.8} or \eqref{eq:4.4}, with the only difference that at each step we might be able to dualize longer column into shorter ones.
    Down to $D=7$, the spectrum contains exactly one graviton, three photons, together with a number of higher $p$-forms and exotic $[p,1]$ tensor. These will be responsible for the additional gravitons upon further compactifying to $D=4$.
    This pattern is general: a $[k,1]$ tensor on a torus $T^{k-1}$ will produce exactly one graviton and $k-1$ photons in dimension $D= k+1$. This is the correct amount to couple to the KK charges, so they could in principle gauge the corresponding symmetry. In lower dimension, the usual  graviphotons will emerge, so we are guaranteed to have enough of them at each step. Let us show this generality, by starting with the origin of the unique graviton in $D=k+3$: it is straightforwardly obtained by removing at each step a box from the long column, without touching the short one. This is in fact the first graviton encountered in dimensional reduction, and it is always alone, for the following reason. Starting from a $[k,1]$-type diagram in dimension $D = 2k+2$, we are guaranteed to meet a graviton after $k-1$ circle reductions, by removing boxes from the long column at each step. A priori, however, we could have encountered a dual graviton in previous steps, via dualizing on the left column, but self-duality in the original representation prevents it. To show this, we want to find out at which step in the reduction a long column would dualize to length 1. The longer a column is the shorter is its dual, so we only need to look at the longest one in the decomposition, which naively is length $k$, obtained by never putting any index in the compact directions. As can be seen in \eqref{eq:4.7} however, already at the first step the self-duality condition leaves the longest representation to have length $k-1$. Let us check after how many steps this representation could dualize to a graviton: after $s$ steps we reach dimension $D-s = 2k+2-s$, so we would need to impose
    \be
        1 = (D-s-2)-(k-1) = k-s+1 \implies s = k
    \ee
    This is one more step then it took earlier to reach the graviton without dualizations, so that graviton is the only one in the spectrum after compactifying on $T^{k-1}$. Notice that it does not matter if some other cancellation happens and reduces the maximum length to, say, $k-2$: this column would take even longer to dualize to a graviton, so the argument still holds.
    
    Let us now count the photons in $D= k+3$. Starting from a $[k,1]$-type diagram, to get photons we need to remove a box from the left column at all steps, and at some point the one in the right column. We are free to do it at any step. Each time it is done, the resulting $p$-form will eventually reduce to a photon on $T^{k-1}$ by taking a single box at any step, therefore giving rise to $k-1$ of them in total, exactly the number one expects from KK reduction.
    
    For the ``box" diagrams of type $[k,k]$ similar arguments lead to quite different behavior. Let us again work through an example first, setting $k=4$.
    \begin{equation}\label{eq:4.10}
        \begin{ytableau}
            \ & \ \\
            \ & \ \\
            \ & \ \\
            \ & \ \\
        \end{ytableau}
        \longmapsto
        \begin{ytableau}
            \ & \ \\
            \ & \ \\
            \ & \ \\
            \ & \ \\
        \end{ytableau}
        \hspace{1mm}\scalebox{1.5}{$\oplus$}\hspace{1mm}
        \begin{ytableau}
            \ & \ \\
            \ & \ \\
            \ & \ \\
            \ \\
        \end{ytableau}
        \hspace{1mm}\scalebox{1.5}{$\oplus$}\hspace{1mm}
        \begin{ytableau}
            \ & \ \\
            \ & \ \\
            \ & \ \\
        \end{ytableau}
        \xrightarrow[\text{duality}]{\text{self}}
        \begin{ytableau}
            \ & \ \\
            \ & \ \\
            \ & \ \\
        \end{ytableau}
    \end{equation}
    Here the double self-duality action, analogously to our guiding example of \eqref{eq:4.3}, identifies three naively different representations, leaving after the first step a single irreducible of $SO(7)$ in $D=9$. On an additional circle, this reduces as
    
    \begin{equation}\label{eq:A.6}
        \begin{ytableau}
            \ & \ \\
            \ & \ \\
            \ & \ \\
        \end{ytableau}
        \longmapsto
        \begin{ytableau}
            \ & \ \\
            \ & \ \\
            \ & \ \\
        \end{ytableau}
        \hspace{1mm}\scalebox{1.5}{$\oplus$}\hspace{1mm}
        \begin{ytableau}
            \ & \ \\
            \ & \ \\
            \ \\
        \end{ytableau}
        \hspace{1mm}\scalebox{1.5}{$\oplus$}\hspace{1mm}
        \begin{ytableau}
            \ & \ \\
            \ & \ \\
        \end{ytableau}   
    \end{equation}
    without any identifications this time.
    \\Going down one more step, we get a copy each of \eqref{eq:A.6} and \eqref{eq:4.3} together with
    \begin{equation}\label{eq:A.7}
        \begin{ytableau}
            \ & \ \\
            \ & \ \\
            \ \\
        \end{ytableau}
        \longmapsto
        \begin{ytableau}
            \ & \ \\
            \ & \ \\
            \ \\
        \end{ytableau}
        \hspace{1mm}\scalebox{1.5}{$\oplus$}\hspace{1mm}
        \begin{ytableau}
            \ & \ \\
            \ & \ \\
        \end{ytableau}
        \hspace{1mm}\scalebox{1.5}{$\oplus$}\hspace{1mm}
        \begin{ytableau}
            \ & \ \\
            \ \\
            \ \\
        \end{ytableau}
        \hspace{1mm}\scalebox{1.5}{$\oplus$}\hspace{1mm}
        \begin{ytableau}
            \ & \ \\
            \ \\
        \end{ytableau}
    \end{equation}
    The only graviton appearing in $D=7$ is the one in \eqref{eq:4.3}, and it is clear that there are no photons. This behaviour is general.
    The reason for the uniqueness of the graviton is precisely the same as in the ``L-shaped" case: the naive way to obtain one by removing at each step a box from both columns is also the fastest way, with the self-duality in the original diagram guaranteeing that we do not hit dual gravitons at any intermediate step. From a $[k,k]$ we then obtain a single graviton by compactifying on $T^{k-1}$. The reason for the absence of photons is also straightforward: to get a single-column object we need to remove all boxes from a column, but this clearly takes at least $k$ steps, and we hit a graviton one step earlier. This shows that theories with an exotic $[k,k]$-form field in dimension $D=2k+2$ have a graviton and a global KK symmetry when compactified on $T^{k-1}$. However, the many additional exotic fields will yield more gravitons under dimensional reduction.  
    \\Finally, we consider the asymmetric $[k,l]$-type diagrams. Again the easiest example shows all the general features, so we set $k=4$, $l=2$ and $D=10$.
    \begin{equation}\label{eq:4.12}
        \begin{ytableau}
            \ & \ \\
            \ & \ \\
            \ \\
            \ \\
        \end{ytableau}
        \longmapsto
        \begin{ytableau}
            \ & \ \\
            \ & \ \\
            \ \\
            \ \\
        \end{ytableau}
        \hspace{1mm}\scalebox{1.5}{$\oplus$}\hspace{1mm}
        \begin{ytableau}
            \ & \ \\
            \ & \ \\
            \ \\
        \end{ytableau}
        \hspace{1mm}\scalebox{1.5}{$\oplus$}\hspace{1mm}
        \begin{ytableau}
            \ & \ \\
            \ \\
            \ \\
            \ \\
        \end{ytableau}
        \hspace{1mm}\scalebox{1.5}{$\oplus$}\hspace{1mm}
        \begin{ytableau}
            \ & \ \\
            \ \\
            \ \\
        \end{ytableau} 
        \xrightarrow[\text{duality}]{\text{self}}
        \begin{ytableau}
            \ & \ \\
            \ & \ \\
            \ \\
        \end{ytableau}
        \hspace{1mm}\scalebox{1.5}{$\oplus$}\hspace{1mm}
        \begin{ytableau}
            \ & \ \\
            \ \\
            \ \\
        \end{ytableau}
    \end{equation}
    At the second step we recover, without any more identifications, the familiar reductions \eqref{eq:4.8} and \eqref{eq:A.7}. Each of these two tensor will produce no gravitons in $D > k+1$ and exactly one in $D = k+3$.
    So in no lower dimensional theory we ever find a single graviton: the moment one appears, at least another one does as well.
    The argument and the general counting goes along the same lines of those above. To obtain a graviton we need to shorten the leftmost column to length 1, so we need $k-1$ circle compactifications. As before, self-duality prevents us to hit dual gravitons along the way. We however also need to shorten the right column to length one, and there are multiple ways to do so: we need to get rid of $l-1$ boxes by erasing either one or zero of them at each step. Seen this way, the counting becomes a  combinatorics problem: the different representations that eventually reduce to a graviton are labeled by different ways to choose the $l-1$ steps in which to remove a box among the $k-1$ available ones. The number of gravitons arising in $D=k+3$ from a generic $[k,l]$ diagram is then
    \be
        N_\text{grav}(k,l) = \binom{k-1}{l-1} = \frac{(k-1)!}{(l-1)!(k-l)!}
    \ee
    which is in agreement with the limiting cases $l=k$ and $l=1$ corresponding to the theories discussed before. We stress again that this counting applies when reducing to dimension $D=k+3$. We already carried out the photon counting explicitly in this dimension, to identify that the KK symmetry in $D=k+3$ could only be gauged in the case of $[k,1]$ diagrams.
    Analogous countings show that for any two column diagrams there are no gravitons in $D > k+3$, and at least two of them in $D<k+3$.

\bibliographystyle{JHEP}
\bibliography{4comma0.bib}

\end{document}